\documentclass[twocolumn,prl,showpacs]{revtex4}
\usepackage{graphicx}
\begin{document}

\newcommand{\re}{\mathop{\mathrm{Re}}}

\newcommand{\be}{\begin{equation}}
\newcommand{\ee}{\end{equation}}
\newcommand{\bea}{\begin{eqnarray}}
\newcommand{\eea}{\end{eqnarray}}


\title{Brane universes tested by supernovae}

\author{Mariusz P. D\c{a}browski}
\email{mpdabfz@uoo.univ.szczecin.pl}
\affiliation{\it Institute of Physics, University of Szczecin, Wielkopolska 15,
          70-451 Szczecin, Poland}
\author{W{\l }odzimierz God{\l }owski}
\email{godlows@oa.uj.edu.pl}
\affiliation{\it Astronomical Observatory, Jagiellonian University, 30-244
Krakow, ul. Orla 171, Poland}
\author{Marek Szyd{\l }owski}
\email{uoszydlo@cyf-kr.edu.pl}
\affiliation{\it Astronomical Observatory, Jagiellonian University, 30-244
Krakow, ul. Orla 171, Poland}

\date{\today}

\input epsf

\begin{abstract}
We discuss observational constrains coming from supernovae Ia \cite{Perlmutter99}
imposed on the behaviour of the Randall-Sundrum models. In the case of dust matter
on the brane, the difference between the best-fit general relativistic model with
a $\Lambda$-term \cite{Perlmutter99} and the best-fit brane models becomes detectable for
redshifts $z > 0.6$.
It is interesting that brane models predict brighter galaxies for such redshifts which
is in agreement with the measurement of the $z = 1.7$ supernova \cite{Riess01}
and with the New Data from the High Z Supernovae Search Team \cite{schmit02}.
We also demonstrate that the fit to supernovae data can also be
obtained, if we admit the "super-negative" dark energy $p = - (4/3) \varrho$ on
the brane, where the dark energy in a way mimics the influence of the cosmological constant.
It also appears that the dark energy enlarges the age of the universe which is
demanded in cosmology. Finally, we propose to check for dark radiation
and brane tension by the application of the angular diameter of galaxies minimum value test.
\end{abstract}

\maketitle

The idea of brane universes has originated from Ho\v{r}ava and
Witten \cite{hw} followed by Randall and Sundrum \cite{rs1+2}.
Brane models admit new parameters which are not present in
standard cosmology (brane tension $\lambda$ and dark radiation ${\cal U}$).
From the astronomical observations of
supernovae Ia \cite{Perlmutter99,Riess98} one knows that the universe
is now accelerating and the best-fit model is for the 4-dimensional cosmological constant
density parameter
$\Omega_{\Lambda_{(4)},0}=0.72$ and for the dust density parameter
$\Omega_{m,0}=0.28$ (index "0" refers to the present moment of time).
In other words, only the exotic (negative pressure) matter in
standard cosmology can lead to this global effect. On the other hand, in
brane models the $\varrho^2$ quadratic contribution in the energy density
even for a small negative pressure, contributes effectively as the
positive pressure, and makes brane
models less accelerating. In this paper we argue that in order to avoid this problem one
requires much stronger negative pressure $p < - \varrho$ matter to be present on the brane (cf. Ref. \cite{darkenergy})
unless the new HZT data shows the $z>1$ supernovae are brighter than expected
\cite{Riess01,schmit02}.

In Ref. \cite{Szydlo02} we gave the formalism to express dynamical equations in
terms of dimensionless observational density parameters $\Omega$.
Following Refs. \cite{Dabrowski96,AJI+II,AJIII}
we introduce the notation useful for this purpose. In this
notation the Friedmann equation for brane universes takes the form
\begin{equation}
\label{FriedCCC}
\frac{1}{a^2} \left( \frac{da}{dt} \right)^2 =
\frac{C_{GR}}{a^{3\gamma}} + \frac{C_{\lambda}}{a^{6\gamma}} -
\frac{k}{a^2} + \frac{\Lambda_{(4)}}{3} + \frac{C_{\cal U}}{a^4} ,
\end{equation}
where $a(t)$ is the scale factor, $k=0,\pm1$ the curvature index, $\Lambda_{(4)}$ the
4-dimensional cosmological contant, and $\gamma$ the barotropic index
($p = (\gamma - 1)\varrho$, $p$ - the pressure, $\varrho$ - the energy
density). In Eq. (\ref{FriedCCC})
we have defined the appropriate constants ($\kappa_{(4)}^2 = 8\pi G$)
$C_{GR} = (1/3)\kappa_{(4)}^2a^{3\gamma} \varrho$,
$C_{\lambda} = 1/6\lambda \cdot \kappa_{(4)}^2 a^{6\gamma}
\varrho^2$, $C_{\cal U} = 2/\kappa_{(4)}^2 \lambda \cdot a^4 {\cal U}$ ,
and $C_{GR}$ is a of general relativistic nature, $C_{\lambda}$ comes as contribution
from brane tension $\lambda$, and $C_{\cal U}$ as a contribution from dark radiation.
Though in Refs. \cite{sopuerta,Szydlo02} the Eq. (\ref{FriedCCC}) was studied using
qualitative methods we just mention here that the cases $\gamma = 0$
(cosmological constant), $\gamma = 1/3$ (domain walls)
and $\gamma = 2/3$ (cosmic strings) can be exactly integrable in terms of elementary
\cite{Dabreg02} or elliptic \cite{Dabrowski96} functions.
For other values of $\gamma = 4/3; 1; 2$, the terms of the type $1/a^8$
and $1/a^{12}$ appear, and the integration involves hyperelliptic
functions. In particular, oscillating non-singular solutions appear for
dark energy $\gamma = -1/3$ \cite{Dabreg02}.

In order to study observational tests we now define
dimensionless observational density parameters \cite{AJI+II,AJIII}
\bea
\label{Omegadef}
\Omega_{GR}  &=&  \frac{\kappa_{(4)}^2}{3H^2} \varrho ,
\hspace{15pt}
\Omega_{\lambda}  =  \frac{\kappa_{(4)}^2}{6H^2\lambda} \varrho^2 ,
\hspace{15pt}
\Omega_{\cal U}  =  \frac{2}{\kappa_{(4)}^2 H^2\lambda} {\cal U}
,\nonumber \\
\Omega_{k}  &=&  - \frac{k}{H^2a^2} ,
\hspace{15pt}
\Omega_{\Lambda_{(4)}}  =  \frac{\Lambda_{(4)}}{3H^2} ,
\eea
where the Hubble parameter $H = \dot{a}/a$, and the deceleration parameter
$q  =  - \ddot{a}a/\dot{a}^2$ ,
so that the Friedmann equation (\ref{FriedCCC}) can be written down
in the form
\begin{equation}
\label{Om=1}
\Omega_{GR} + \Omega_{\lambda} + \Omega_{k} + \Omega_{\Lambda_{(4)}} + \Omega_{\cal U}
= 1  .
\end{equation}
Note that $\Omega_{\cal U}$ in (\ref{Omegadef}), despite standard radiation term, can either be
positive or negative.
Using (\ref{Omegadef}), the equation (\ref{FriedCCC}) can now be
rewritten as (compare Eq.(10) of \cite{AJI+II})
\begin{equation}
\label{Lambda4}
\Omega_{\Lambda_{(4)}}  = \frac{3\gamma - 2}{2} \Omega_{GR} +
(3\gamma - 1) \Omega_{\lambda} + \Omega_{\cal U} - q .
\end{equation}
It is also useful to express the curvature of spatial sections
by observational parameters by using (\ref{Om=1}) and (\ref{Lambda4})
\begin{equation}
\label{indexk}
- \Omega_{k} = \frac{3\gamma}{2} \Omega_{GR} +
3\gamma \Omega_{\lambda} + 2\Omega_{\cal U} - q - 1.
\end{equation}
These relations (\ref{Lambda4}) and (\ref{indexk}) allow to write down an explicit
redshift-magnitude relation (a generalized Hubble law) for the brane models to study
their compatibility with astronomical data which
is the subject of the present paper. Obviously, the luminosity of galaxies depends
on the present densities of the different components
of matter content $\Omega$ given by (\ref{Omegadef}) and their equations of state reflected
by the value of the barotropic index $\gamma$.

Let us consider an observer located at $r=0$ at the moment $t=t_0$
which receives a light ray emitted at $t=t_1$ from the source of the absolute luminosity $L$
located at the radial distance $r_1$. The redshift z
of the source is related to the scale factor $a(t)$ at the two moments of evolution
by $1+z=a(t_0)/a(t_1) \equiv a_0/a$.
If the apparent luminosity of the source as measured by the observer is $l$,
then the luminosity distance $d_L$ of the source is defined by the
relation
\be
\label{luminosity}
l={L\over 4\pi d_L^2},
\ee
where
\be
\label{DeeL}
d_L=(1+z)a_0 r_1 \equiv \frac{{\cal D}_L(z)}{H_0},
\ee
and ${\cal D}_L$ is the dimensionless luminosity distance.
The observed and absolute luminosities are
defined in terms of K-corrected apparent and absolute
magnitudes $m$ and $M$. When written in terms of $m$ and $M$, Eq.(\ref{luminosity}) yields
\be
\label{m(z)}
m(z)={\cal M} + 5\log_{10}
[{\cal D}_L(z)],
\ee
where ${\cal M}=M-5\log_{10}H_0+25$.
For homogeneous and isotropic Friedmann models one gets \cite{AJIII}
\be
\label{Deelfin}
{\cal D}_L(z) = \frac{\left( 1+z \right)}
{\sqrt{{\cal K}}} S(\chi)
\ee
where $S(\chi)=\sin \chi$ for ${\cal K}=-\Omega_{k,0}$; $S(\chi)=\chi $
for ${\cal K}=1$; $S(\chi)=\sinh $ for ${\cal K}=\Omega_{k,0}$.
From the Friedmann equation (\ref{FriedCCC}) and the form of the FRW metric
we have
\bea
\label{chir1}
\chi(z)={1\over a_0 H_0}\int\limits_0^z
\left\{\Omega_{\lambda,0}\left(1+z^{'} \right)^{6\gamma} +
\Omega_{GR,0}\left(1+z^{'} \right)^{3\gamma}
\right. \nonumber \\ \left. + \Omega_{k,0}\left(1+z^{'} \right)^2 +
\Omega_{{\cal U},0}\left(1+z^{'}\right)^4 +
\Omega_{\Lambda_{(4),0}}\right\}^{-1/2}dz^{'}.
\eea
Firstly, we will study the case $\gamma = 1$ (dust on the brane;
we will label $\Omega_{GR}$ by $\Omega_m $).
The case $\gamma = 2/3$
(cosmic strings on the brane) has recently been studied in \cite{Singh1/3} where, in fact,
$\Omega_{\cal U}$ and $\Omega_{\lambda}$ were neglected and
where the term $\Omega_{m,0} (1+z^{'})^3$ was
introduced in order to admit dust matter on the brane. This case was already presented in
a different framework in Ref. \cite{AJIII}.
Secondly, we will study the case $\gamma = -1/3$ (dark energy on the brane \cite{darkenergy}
- we will label this type of matter with $\Omega_{d}$ instead of $\Omega_{GR}$).

Now we test brane models using the sample of Ref. \cite{Perlmutter99}.
In order to avoid any possible selection effects, we use the full sample
(usually, one excludes
two data points as outliers and another two points, presumably reddened,
from the full sample of 60 supernovae). It means that our basic sample is
the sample A of Ref. \cite{Perlmutter99}.
We test our model using the likelihood method \cite{Riess98}.

First of all, we estimated the value of ${\cal M}$ from the sample
of 18 low redshift supernovae, also testing our result by the full sample
of 60 supernovae taking $\Omega_{\lambda}=0$.
We obtained ${\cal M}=-3.39$ which is in a very good agreement with the
results of Refs. \cite{Efstathiou99,Vishwakarma}.
Also, we obtained for the model of Ref. \cite{Perlmutter99} same value of $\chi^2 =
96.5$.

Neglecting dark radiation $\Omega_{{\cal U},0} = 0$ we formally got
the best fit ($\chi^2=94.6$) for $\Omega_{k,0}=-0.9$,
$\Omega_{m,0}=0.59$, $\Omega_{\lambda,0}=0.04$, $\Omega_{\Lambda,0}=1.27$,
which is completely unrealistic, because $\Omega_{m,0}=0.59$ is too large in
comparison with the observational limit (also $\Omega_{k,0}=-0.9$
is not very realistic from the observational point of view).

\begin{figure}[h]
\includegraphics[angle=270,scale=.31]{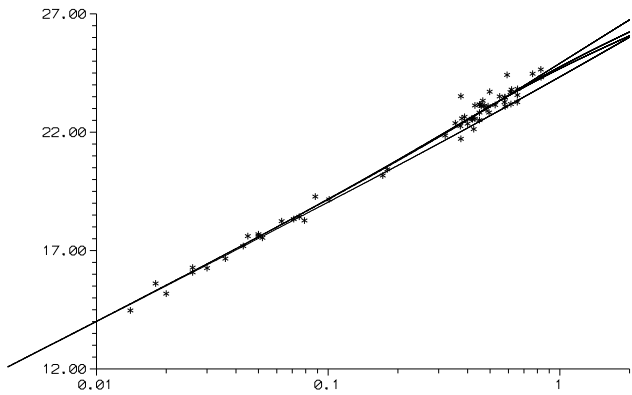}
\caption{Redshift-magnitude relations for $\gamma = 1$ brane universes (dust on the brane).
The top line is the best-fit flat model of Ref. \cite{Perlmutter99} with $\Omega_{m,0}=0.28$,
$\Omega_{\Lambda_{(4)},0}=0.72$. The bottom line is a pure flat model with
$\Omega_{\Lambda_{(4)},0}=0.$ Between these two there are brane models
with $\Omega_{\lambda,0} \ne 0$: lower - the best-fit
model; higher - the best-fit {\it flat} model.}
\label{Fig.1}
\end{figure}

However, we should note that, in fact, we have
an ellipsoid of admissible models in a 3D parameter space $\Omega_{m,0}$,
$\Omega_{\lambda,0}$, $\Omega_{\Lambda_{(4)},0}$ at hand. Then, we have more freedom
than in the case of analysis of Ref. \cite{Perlmutter99} where they had only an ellipse
in a 2D parameter space $\Omega_{m,0}$, $\Omega_{\Lambda_{(4)},0}$.
For a flat model $\Omega_{k,0}=0$ we obtain "corridors" of possible models (\ref{Fig.1}).
Formally, the best-fit flat model is $\Omega_{m,0}=0.01$, $\Omega_{\lambda,0}=0.09$
$\chi^2=94.7$ which is again unrealistic.
In the realistic case we obtain for a flat model $\Omega_{m,0}=0.25$,
$\Omega_{\lambda,0}=0.02$, $\Omega_{\Lambda_{(4)},0}=0.73$
with $\chi^2=95.6$.
One should note that all realistic brane models require also the presence of the
positive 4-dimensional cosmological constant ($\Omega_{\Lambda_{(4)},0} \sim
0.7$).

There is a question if we could fit a model with negative
$\Omega_{\lambda,0}$? For instance, in a flat Universe we could
fit the model with $\Omega_{m,0}=0.35$ (too much in comparison
with the observational limits on the mass of the cluster of galaxies)
$\Omega_{\lambda,0}=-0.01$, $\Omega_{\Lambda,0}=0.66$ ($\chi^2=96.3$).
However, it is not possible to fit any models
with more negative values of $\Omega_{\lambda,0}$,  regardless the Universe
is flat or not.

In Fig.\ref{Fig.1} we present plots of redshift-magnitude relations against
the supernovae data.
One can observe that in both cases (best-fit and best-fit flat models) the
difference between brane models and a pure flat
model with $\Omega_{\Lambda_{(4)},0}=0$ is largest for $0.6 < z < 0.7$
while it significantly decreases for the higher redshifts. It gives us a possibility
to discriminate between
general relativistic models and brane models when the data from high-redshift
supernovae $z>1$ is available.On the other hand,
the difference between the best-fit general relativistic model with
a $\Lambda$-term \cite{Perlmutter99} and the best-fit brane models becomes detectable for
redshifts $z > 0.6$.
It is interesting that brane models predict brighter galaxies for such redshifts which
is in agreement with the measurement of the $z = 1.7$ supernova \cite{Riess01}
and with the New Data from the High Z Supernovae Search Team \cite{schmit02}.
In other words, if the farthest $z >1$ supernovae were brighter, the brane universes
would be the reality.

One should note that we made our analysis without excluding any supernovae from the sample.
However, from the formal point of view, when we analyze the full sample A,
all models should be rejected even on the confidence level of 0.99.
One of the reason could be the fact that the assumed error bars are
too small. However, in majority of papers another solution is suggested.
Usually, one excludes 2 supernovae as outliers, and 2 as reddened from the sample
of 42 high-redshift supernovae and eventually 2 outliers from the sample of 18
low-redshift supernovae.
We decided to use the full sample A as our basic sample
because a rejection of any supernovae from the sample can be the source
of lack of control for selection effects. However, for completness, we also
made our analysis using samples B and C.
It emerged that it does not significantly changes our results, though,
increases quality of the fit.
Formally, the best fit for the sample B is (56 supernovae)
($\chi^2=57.3$): $\Omega_{k,0}=-0.1$
$\Omega_{m,0}=0.17$, $\Omega_{\lambda,0}=0.06$, $\Omega_{\Lambda_{(4)},0}=0.87$.
For the flat model we obtain ($\chi^2=57.3$):
$\Omega_{m,0}=0.12$, $\Omega_{\lambda,0}=0.06$,
$\Omega_{\Lambda_{(4)},0}=0.82$,
while for a "realistic" model ($\Omega_{m,0}=0.25$, $\Omega_{\lambda,0}=0.02$)
$\chi^2=57.6$.
Formally, the best fit for the sample C (54 supernovae)
($\chi^2=53.5$) gives $\Omega_{k,0}=0$ (flat)
$\Omega_{m,0}=0.21$, $\Omega_{\lambda,0}=0.04$,
$\Omega_{\Lambda_{(4)},0}=0.75$,
while for "realistic" model ($\Omega_{m,0}=0.27$, $\Omega_{\lambda,0}=0.02$)
$\chi^2=53.6$.

One should note that we have also separately estimated the value of ${\cal M}$
for the sample B and C.
We obtained ${\cal M}=-3.42$ which is again in a good agreement with
the results of Ref. \cite{Efstathiou99} (for a "combined" sample one
obtains ${\cal M}=-3.45$). However, if we use this value in our analysis
it does not change significantly the results ($\chi^2$ does not change more than
$1$ which is a marginal effect for $\chi^2$ distribution for 53 or 55 degrees
of freedom).

\begin{figure}[h]
\includegraphics[angle=270,scale=.31]{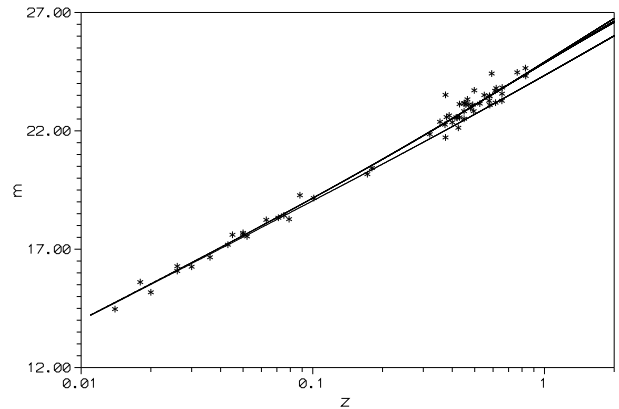}
\caption{A redshift-magnitude relation for $\gamma = -1/3$ brane universes
(dark energy on the brane). The top and bottom lines are same as in Fig. \ref{Fig.1}.
The brane dark energy models plots are very close to the top line of Ref. \cite{Perlmutter99}.}
\label{Fig.2}
\end{figure}

In Fig. \ref{Fig.2} we present a redshift-magnitude relation (\ref{chir1}) for
brane models with dark energy ($\gamma =
-1/3$). Note that the theoretical curves are very close to that of
\cite{Perlmutter99} which means that the dark energy
cancels the positive-pressure influence of the $\varrho^2$ term
and can simulate the negative-pressure influence of the cosmological constant to cause
cosmic acceleration. From the formal point of view the best fit
is ($\chi^2=95.4$) for $\Omega_{k,0}= 0.2$, $\Omega_{d,0}=0.7$, $\Omega_{\lambda,0}=-0.1$,
$\Omega_{\cal U} = 0.2$, $\Omega_{\Lambda_{(4)},0}=0$ which
means that the cosmological constant must necessarily {\it vanish}. From
this result we can conclude that the dark energy $p = - (4/3) \varrho$ can {\it
mimic} the contribution from the $\Lambda_{(4)}$-term in standard models.

For the best-fit flat model ($\Omega_{k,0}=0$) we have
($\chi^2=95.4$): $\Omega_{d,0}=0.2$, $\Omega_{\lambda,0}=-0.1$,
$\Omega_{\cal U} = 0.2$, $\Omega_{\Lambda_{(4)},0}=0.7$.

\begin{figure}[h]
\includegraphics[angle=270,scale=.31]{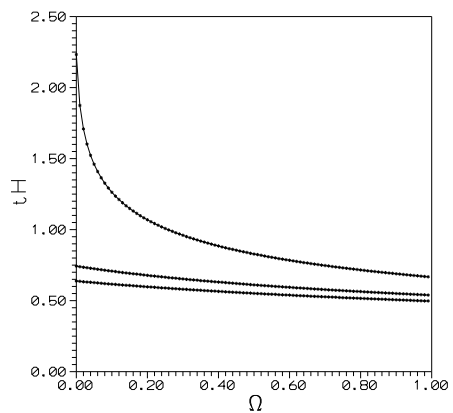}
\caption{The age of the universe $t_0$ in units of $H_0^{-1}$ for the brane models with dust ($0 \leq \Omega_{m,0} \leq 1$
on the horizontal axis). Here $\Omega_{{\cal U},0} = \Omega_{k,0} = 0$,
$\Omega_{{\lambda},0} = 0, 0.05, 0.1$ (top, middle, bottom).}
\label{Fig.3}
\end{figure}

\begin{figure}[h]
\includegraphics[angle=270,scale=.31]{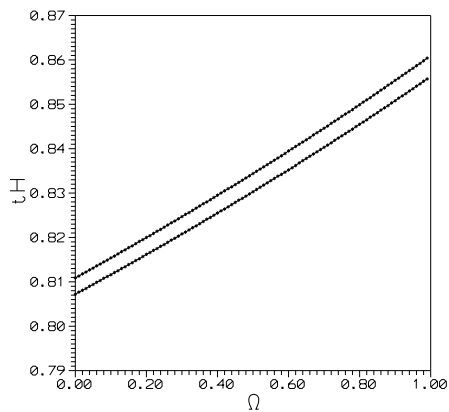}
\caption{The age of the universe $t_0$ in units of $H_0^{-1}$ for the brane models with
dark energy on the brane
($0 \leq \Omega_{d,0} \leq 1$ on the horizontal axis). Here $\Omega_{{\cal U},0} =
0.2$, $\Omega_{{\lambda},0} = 0.05, 0$ (top, bottom) which shows weaker influence
of the brane effects to increase the age.}
\label{Fig.4}
\end{figure}

Now let us briefly discuss the effect of brane parameters and dark energy onto
the age of the universe which according to (\ref{FriedCCC}) is given by
\bea
\label{age}
H_0 t_0 = \int_0^1 \left\{ \Omega_{GR,0} x^{-3\gamma +
4} + \Omega_{{\lambda},0} x^{-6\gamma + 4}
\right. \nonumber \\ \left.
+ \Omega_{{\cal U},0} +
\Omega_{k,0} x^2 + \Omega_{\Lambda_{(4)},0} x^4 \right\}^{-\frac{1}{2}} x dx  ,
\eea
where $x = a/a_0$. We made a plot for the dust $\gamma = 1$ on the brane in
Fig.\ref{Fig.3} which shows that the effect of quadratic in
energy density term represented by $\Omega_{\lambda}$ is to {\it
lower} significantly the age of the universe. The problem can be
avoided, if we accept the dark energy $\gamma = - 1/3$ \cite{darkenergy}
on the brane, since the dark energy
has a very strong influence to increase the age \cite{Dabreg02}.
In Fig. \ref{Fig.4} we made a plot for this case which shows how
the dark energy enlarges the age.

Finally, let us study the angular diameter test for brane universes.
The angular diameter of a galaxy is defined by
\be
\label{angdiam}
\theta = \frac{d(z+1)^2}{d_{L}} ,
\ee
where $d$ is a linear size of the galaxy. In a flat dust ($\gamma
=1$) universe $\theta$ has the minimum value $z_{min} = 5/4$.
It is particularly interesting to notice that
for flat brane models with $\Omega_{\lambda} \approx 0, \Omega_{\Lambda_{(4)}}
\approx 0$ the dark radiation can {\it enlarge} the minimum value of $\theta$
while the ordinary radiation lowers this value \cite{AJI+II}, i.e.,
\be
\label{zminU}
z_{min} = \frac{1}{2{\cal U}} \left( \Omega_{\cal U} - 1 + \sqrt{3
\Omega_{\cal U} + 1} \right) \ge \frac{5}{4}
\ee
for $\Omega_{\cal U} \leq 0$. This is a general influence of negative dark radiation
onto the angular diameter size for brane models. One can also notice that there
exists a restriction on the amount of negative dark radiation coming from
(\ref{zminU}) $(\Omega_{\cal U} \geq - 1/3)$ which can serve
as a test for the admissible value of $\Omega_{\cal U} = - 1/3$ ($z_{min}
=2$) in order to observe the minimum.

\begin{figure}[h]
\includegraphics[scale=.31]{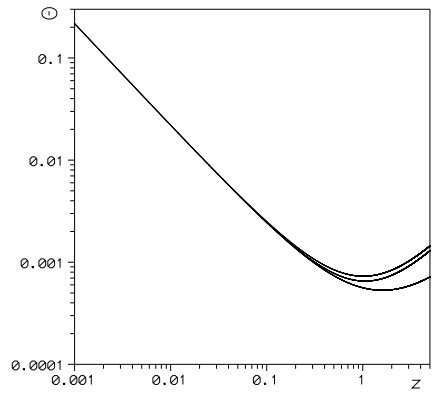}
\caption{The angular diameter $\theta$ for $\Omega_{\lambda} = 0.1, \Omega_m = 0.3,
\Omega_{\Lambda_{(4)}} = 0.72$, and the two values of $\Omega_{\cal U} = 0.1, -0.1$
(top, middle) in comparison with the model of Ref. \cite{Perlmutter99} with $\Omega_m = 0.28,
\Omega_{\Lambda_{(4)}} = 0.72$ (bottom).}
\label{Fig.5}
\end{figure}

More detailed analytic and numerical studies \cite{Dabreg02} show
that the increase of $z_{min}$ is even more sensitive to negative
values of $\Omega_{\lambda}$ ($\varrho^2$ contribution). Similarly as for the dark
radiation $\Omega_{\cal U}$, the minimum disappears for some large negative $\Omega_{\lambda}$.
Positive $\Omega_{\cal U}$ and $\Omega_{\lambda}$ make $z_{min}$ decrease.
In Fig. \ref{Fig.5} we present a plot from which one can see the
sensitivity of $z_{min}$ to $\Omega_{\cal U}$. We have also checked
\cite{Dabreg02} that the dark energy $\Omega_d$ has very little influence
onto the value of $z_{min}$.


\begin{thebibliography}{99}

\bibitem{hw} P. Ho\v{r}ava and E. Witten, Nucl. Phys. B{\bf 460}
(1996), 506; {\it ibid} B{\bf 475}, 94.

\bibitem{rs1+2} L. Randall and R. Sundrum, Phys. Rev. Lett., {\bf 83}
(1999), 3370; {\it ibid} {\bf 83} (1999), 4690.

\bibitem{Perlmutter99}  S. Perlmutter {\it et al.}, Ap. J. {\bf 517},
(1999) 565.

\bibitem{Riess98} A. G. Riess {\it et al.} Astron. J. {\bf 116}
(1998) 1009.

\bibitem{darkenergy} R.R. Caldwell, astro-ph/9908168,
S. Hannestad and E. M\"orstell, astro-ph/0205096, P.H. Frampton, astro-ph/0209037.

\bibitem{Riess01} A.G. Riess et al., Ap. J. {\bf 560} (2001), 49.

\bibitem{schmit02} B.P. Schmidt {\it et al.} - private communication.

\bibitem{Szydlo02} M. Szyd{\l }owski, M.P. D\c{a}browski, and A.
Krawiec, Phys. Rev. D{\bf 66} (2002), 0640XX (hep-th/0201066).

\bibitem{Dabrowski96} M.P. D\c{a}browski, Ann. Phys (N. Y.) {\bf 248} (1996) 199.

\bibitem{AJI+II} M.P. D\c{a}browski and J. Stelmach, Astron. Journ. {\bf 92}
(1986), 1272; {\it ibid} {\bf 93} (1987), 1373.

\bibitem{AJIII} M.P. D\c{a}browski and J. Stelmach, Astron. Journ. {\bf 97}
(1989), 978.

\bibitem{sopuerta} A. Campos and C.F. Sopuerta, Phys. Rev. D{\bf 63}, 104012 (2001).

\bibitem{Efstathiou99} G. Efstathiou, Mon. Not. R. Astr. Soc. {\bf 303}
(1999), 147.

\bibitem{Vishwakarma} Vishwakarma, Gen. Rel. Grav. {\bf 33} (2001),
1973.


\bibitem{Singh1/3} P. Singh, R.G. Vishwakarma and N. Dadhich, hep-th/0206193.

\bibitem{Dabreg02} M.P. D\c{a}browski, W. God{\l }owski, and M.
Szyd{\l }owski, in preparation.

\end{thebibliography}
\end{document}